# Synthetic Trust Attacks:

## Modeling How Generative AI Manipulates Human Decisions in Social Engineering Fraud

**Muhammad Tahir Ashraf (BeyondTahir)**

Chair, AAAI Pakistan Chapter (aaai.pk) — Chapter ID: 647885
Founder and CEO, PureDesigners Ltd. | IBM Watson Silver Partner
ORCID: 0009-0001-3400-5016 | research@beyondtahir.com

## Abstract

Imagine receiving a video call from your CFO, surrounded by colleagues, asking you to urgently authorize a confidential transfer. You comply. Every person on that call was fake, and you just lost $25 million. This is not a hypothetical. It happened in Hong Kong in January 2024 [7], and it is becoming the template for a new generation of fraud. AI has not invented a new crime. It has industrialized an ancient one: the manufacture of trust. Attackers today no longer need to break into systems; they manufacture credibility so convincing that victims authorize the breach themselves. In 2024, the FBI's Internet Crime Complaint Center recorded 859,532 fraud complaints with $16.6 billion in reported losses [1]. World Economic Forum surveys confirm that majorities of executives are experiencing increases in cyber-enabled fraud at organizational and personal levels [2]. Generative AI is the force-multiplier behind this surge, enabling voice cloning, video deepfakes, and LLM-driven conversations that scale social engineering attacks to industrial speed.

This paper proposes Synthetic Trust Attacks (STAs) as a formal threat category and introduces STAM (the Synthetic Trust Attack Model), an eight-stage operational framework covering the full attack chain from adversary reconnaissance through post-compliance leverage. The core argument is this: existing defenses target synthetic media detection, but the real attack surface is the victim's decision. When human deepfake detection accuracy sits at approximately 55.5%, barely above chance [10], and LLM scam agents achieve 46% compliance versus 18% for human operators while evading safety filters entirely [5], the perception layer has already failed. Defense must move to the decision layer. To support this reframing, we present a five-category Trust-Cue Taxonomy, a reproducible 17-field Incident Coding Schema with a pilot-coded example, and four falsifiable hypotheses linking attack structure to compliance outcomes. The paper further operationalizes the author's practitioner-developed Calm, Check, Confirm protocol as a research-grade decision-layer defense. Synthetic credibility, not synthetic media, is the true attack surface of the AI fraud era.

# 1. Introduction

The sociology of deception has not changed. What has changed is the industrial capacity to execute deception at scale. Generative AI does not invent fraud; it perfects it, enabling attackers to construct highly personalized, multimodal social engineering attacks that compress verification windows, manufacture authority, and exploit cognitive heuristics faster than institutional defenses can adapt [11]. The result is a qualitative transformation in the fraud threat landscape: the emergence of what this paper terms Synthetic Trust Attacks.

Existing academic literature on AI-enabled social engineering has largely concentrated on three capability pillars: (i) the realism of AI-generated content; (ii) the personalization of targeting through large-scale data harvesting; and (iii) the automation of attack infrastructure, enabling mass-scale spear-phishing and vishing campaigns [11][3]. These are genuine contributions. However, they share a critical blind spot: they model the attacker's capabilities without modeling the victim's decision process. The most operationally decisive question in fraud remains undertheorized.

The January 2024 Hong Kong case makes this gap vivid. A finance professional attended a group video conference with what appeared to be multiple familiar colleagues, and authorized transfers totaling approximately HK$200 million. The conference was constructed from pre-recorded deepfake media. The failure was not a failure of awareness; it was a failure of decision architecture under conditions of manufactured authority, contextual plausibility, social proof, and compressed verification time [7].

This paper makes five contributions. First, we formally define Synthetic Trust Attacks as a distinct threat category. Second, we introduce STAM, an eight-stage codeable framework. Third, we present a five-category Trust-Cue Taxonomy bridging computer science, criminology, and social psychology. Fourth, we propose a reproducible Incident Coding Schema. Fifth, we operationalize the author's practitioner-developed Calm, Check, Confirm defensive framework as falsifiable research hypotheses.

# 2. Related Work

## 2.1 Generative AI as a Social Engineering Amplifier

The systematic literature has converged on a three-pillar amplification model [11]. Content realism refers to the capacity of generative models to produce text, audio, and video indistinguishable from authentic human output. Targeting and personalization encompasses LLM-driven bespoke pretexting. Automated attack infrastructure includes scalable AI agents sustaining extended social engineering conversations. This framework maps attacker capabilities but does not explain compliance, the actual outcome that determines fraud success.

## 2.2 Deepfake Threats and Human Detection Limits

The Europol Innovation Lab characterizes deepfakes as a significant threat to law enforcement through the 2030 horizon [3]. INTERPOL's Asia-Pacific assessment documents an alarming rise in AI-enabled deepfake scams and industrial-scale fraud operations [4]. A systematic meta-analysis reports aggregate human deepfake detection accuracy of approximately 55.5%, with confidence intervals approaching chance levels under time pressure [10]. If detection is not feasible in real time, then detection-based defenses alone are structurally insufficient.

## 2.3 LLM Persuasion and the Compliance Evidence

The most directly relevant empirical finding is the study 'Love, Lies, and Language Models' [5]. An LLM-powered scam agent achieved compliance rates of 46% compared to 18% for human operators across extended week-long conversations, while popular safety filters detected 0.0% of the romance-baiting dialogues. This is direct empirical evidence that generative AI already outperforms human operators in sustained trust-building and compliance extraction.

### 2.4 Provenance, Trust Side Effects, and the Decision-Layer Gap

NIST documents trust side effects arising from content provenance labeling [6]. When provenance is marked falsified or incomplete, users distrust even authentic content more strongly than when no provenance signal appears. No existing framework models the complete chain from attacker capabilities to victim compliance decision at an operational level supporting reproducible empirical analysis. STAM is designed to fill this gap.

## 3. Threat Model and Definitions

### 3.1 Formal Definition

A Synthetic Trust Attack (STA) is a social engineering campaign in which an adversary employs generative AI to manufacture, bundle, and deliver trust cues including identity simulation, contextual plausibility signals, authority indicators, and linguistic style mimicry, through orchestrated multi-channel interactions, activating psychological compliance triggers to compress the victim's verification window and induce a target action (payment authorization, credential disclosure, or sensitive data release) under conditions of manufactured trust.

### 3.2 Attacker and Victim Models

The threat actor possesses: access to open-source intelligence on the target; access to voice cloning, video deepfake generation, and LLM-driven conversation tools; capability to operate across multiple communication channels; and organizational infrastructure for multi-day operations including transnational concealment of proceeds. The victim model does not assume naivety. Vulnerability derives from cognitive constraints under artificial urgency and authority pressure. Dual-process theory predicts that under time pressure and emotional loading, human decision-making shifts from deliberative to automatic processing [14], increasing susceptibility to authority and social-proof heuristics. STAs systematically exploit this shift.

## 4. STAM: The Synthetic Trust Attack Model

STAM formalizes the operational chain of a Synthetic Trust Attack through eight sequentially dependent but iteratively reinforcing stages. The model is designed to be codeable: each stage maps to specific observable variables extractable from incident reports, law enforcement advisories, and victim accounts. Figure 1 illustrates the full eight-stage chain.

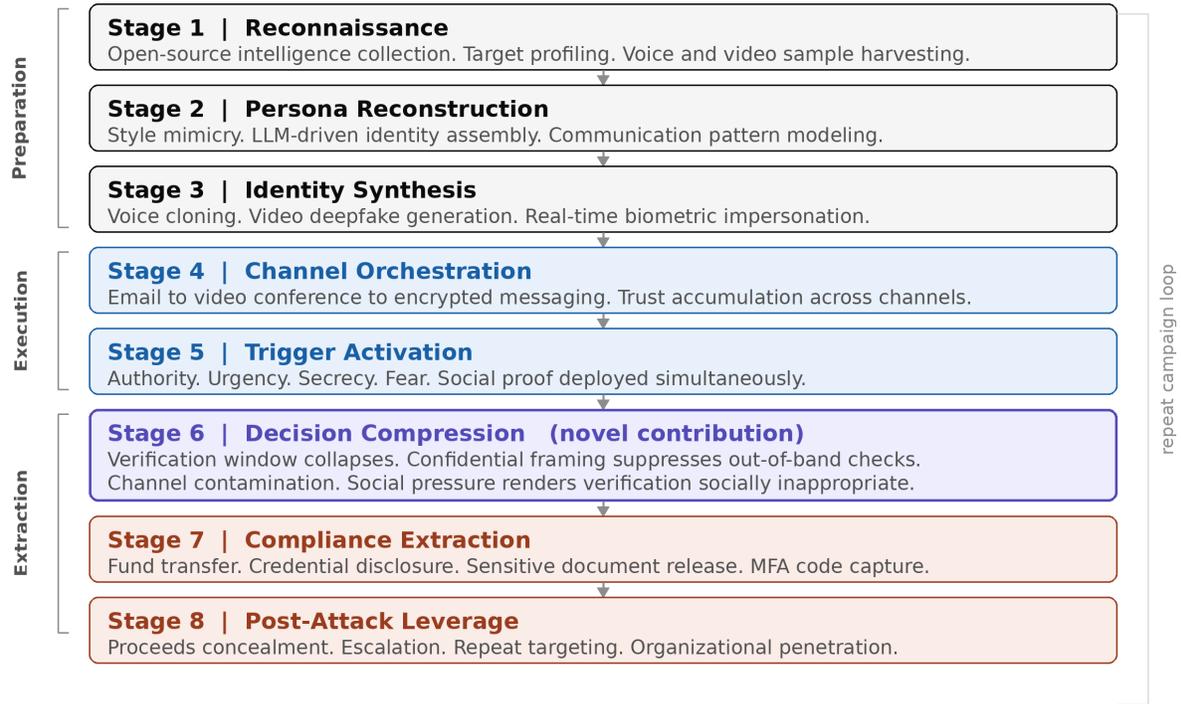

Figure 1. STAM: Synthetic Trust Attack Model. Eight operational stages from reconnaissance to post-attack leverage. Stage 6 (Decision Compression) is the novel contribution of this paper. The feedback loop indicates adversaries may iterate within stages before proceeding, and repeat campaigns against the same target or network.

### 4.1 Stage 1: Reconnaissance

Reconnaissance encompasses systematic open-source intelligence collection on the target and their organizational environment, including harvesting of voice samples, video footage, communication style artifacts, and relational network maps necessary for identity synthesis in Stage 3 [11].

### 4.2 Stage 2: Persona Reconstruction

Persona reconstruction involves assembling the behavioral and communicative profile of the identity to be impersonated. LLM fine-tuning or in-context style mimicry enables generation of communications indistinguishable from the persona's authentic output [5].

### 4.3 Stage 3: Identity Synthesis

Identity synthesis is the technical production of synthetic biometric-like cues, including voice cloning, video deepfake generation, and real-time audio and video manipulation. Real-time deepfake systems capable of rendering interactive face and voice synthesis during a live call represent the current frontier. Human detection accuracy of approximately 55.5% [10] means synthesis does not need to be perfect; it only needs to be plausible under emotional pressure and time constraint.

### 4.4 Stage 4: Channel Orchestration

Channel orchestration describes the deliberate sequencing of communication channels to maximize trust accumulation and minimize verification opportunity. A canonical STA sequence is email establishing context, followed by video conference delivering synthetic presence and social proof, followed by encrypted messaging extracting compliance outside organizational logging. The Hong Kong CFO case followed precisely this pattern [7].

### 4.5 Stage 5: Trigger Activation

Trigger activation involves calibrated deployment of psychological compliance triggers aligned with Cialdini's influence principles [14]. In the STA context: authority is delivered through a simulated senior executive; urgency through time-critical transaction framing; secrecy through instructions not to discuss the matter; social proof through multiple apparent conference participants; and fear through implied reputational or legal consequences of non-compliance.

### 4.6 Stage 6: Decision Compression (Novel Contribution)

Decision compression is the core novel contribution of this paper's threat model. It refers to the deliberate reduction of the victim's effective verification window: the time and channel space available for out-of-band identity confirmation before a compliance action is required. Decision compression is operationalized through explicit time framing, authority framing, secrecy framing, and channel contamination, ensuring all accessible communication routes are occupied by the attack rather than by verification resources. In the Hong Kong CFO case, the confidential transaction framing combined with a live conference-call format created conditions in which interrupting the call to verify participants' identities would have appeared professionally inappropriate [7]. Decision window duration is a codeable variable in the STAM schema.

### 4.7 Stage 7: Compliance Extraction

Compliance extraction is the requested action: fund transfer, credential disclosure, MFA code capture, or sensitive document release. IC3 reporting classifies the dominant categories in dollar terms as business email compromise, investment fraud, and personal data compromise [1].

## 4.8 Stage 8: Post-Attack Leverage

Post-attack leverage encompasses proceeds concealment through multi-layered financial routing and cryptocurrency conversion, escalation through follow-on requests exploiting established compliance precedent, repetition targeting the same victim or network, and organizational penetration using obtained credentials. INTERPOL documents the transnational infrastructure supporting these operations [4].

## 5. Trust-Cue Taxonomy

The Trust-Cue Taxonomy organizes the components of synthetic credibility into five measurable categories. Figure 2 presents the full taxonomy. Each cue category maps to specific STAM stages and specific coding variables in the Incident Schema.

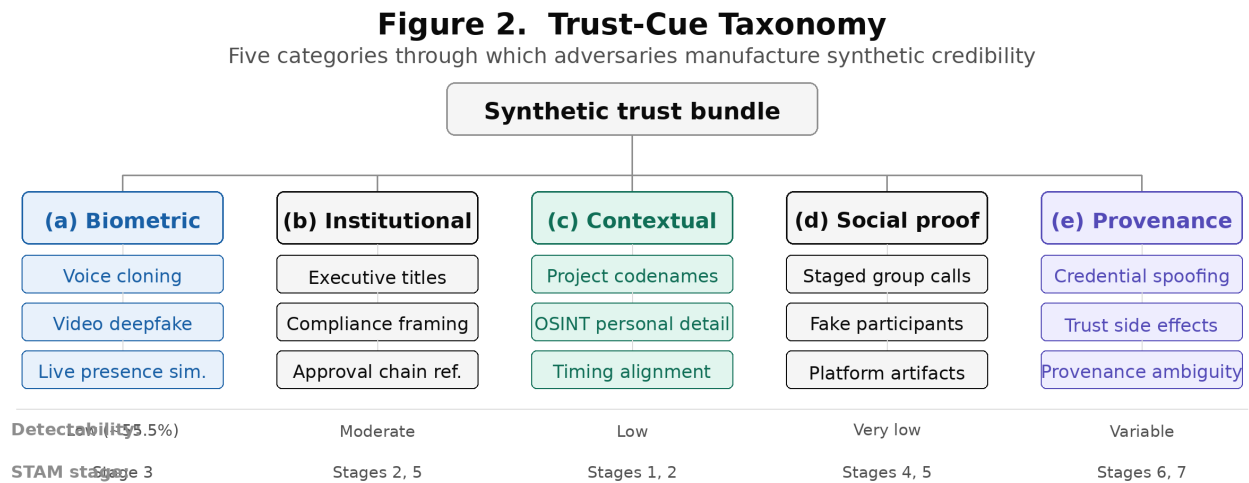

Figure 2. Trust-Cue Taxonomy. Five categories through which adversaries manufacture synthetic credibility, with sub-types, detectability ratings, and corresponding STAM stages. All five categories were simultaneously active in the Hong Kong CFO deepfake incident (January 2024).

### 5.1 Biometric-like Cues

Biometric-like cues exploit the human tendency to treat familiar voice and face recognition as identity verification. Because aggregate human detection accuracy is approximately 55.5% [10] and is likely lower under time pressure and emotional loading, these cues represent the highest-impact single category. The maturation of real-time voice cloning and video synthesis elevates this from a staged-media threat to an interactive one [12].

### 5.2 Institutional Cues

Institutional cues leverage deference to apparent organizational authority through correct executive titles, reference to known approval workflows, compliance and legal language, and document artifacts mirroring internal formatting. Institutional cues become significantly more compelling when combined with contextual cues providing correct internal project terminology and timing. Both were decisive in the Hong Kong CFO case [7].

### 5.3 Contextual Cues

Contextual cues make a request feel internal by referencing the target's current projects, organizational relationships, and professional schedule. Generative AI enables adversaries to personalize pretexts at scales previously requiring large human teams, drawing on OSINT data [11].

### 5.4 Social-Proof Cues

Social-proof cues exploit the heuristic that if multiple trusted peers accept a situation as legitimate, it likely is. In STAs, this is operationalized through staged multi-participant video conferences in which multiple attendees are synthetic. Apparent consensus among familiar colleagues powerfully suppresses individual skepticism, as documented in the Hong Kong case [7] and FBI advisories [8].

### 5.5 Provenance Cues

Provenance cues represent the emerging frontier. As organizations adopt content authenticity standards including C2PA content credentials [13], adversaries gain incentive to spoof or corrupt provenance signals. NIST's documented trust side effects show that falsified provenance can reduce trust even in authentic content [6]. Adversaries who introduce provenance ambiguity into an organization's information environment can degrade the reliability of authenticity signals as a verification resource.

## 6. Incident Coding Schema and Dataset Design

The STAM Incident Coding Schema provides a reproducible framework for analyzing synthetic trust attacks across jurisdictions, victim segments, and time periods. Table 1 presents the full schema. Table 2 presents pilot coding of the Hong Kong CFO case (2024) across all schema fields.

| Field | Values / Format | Coding Notes |
|---|---|---|
| incident_id | Sequential alphanumeric | Format: ISO Country, Year, Number |
| date_range | YYYY-MM to YYYY-MM | Use incident date range; source publication date if unavailable |
| jurisdiction | ISO 3166-1 alpha-2 | Country where fraud occurred; multi-value if transnational |
| victim_segment | Enterprise, Consumer, Government, Mixed | Based on primary victim description |
| impersonated_identity | Executive, Colleague, Family, Authority, Vendor, Unknown | Most senior identity simulated |

| Field | Values / Format | Coding Notes |
|---|---|---|
| channel_sequence | Ordered list: Email, Phone, Video, Messaging, In-person | All channels in temporal order |
| channel_switch_count | Integer 0 or greater | Number of channel transitions |
| modality_set | Text, Audio, Video, Image, Multimodal | All modalities present |
| trigger_tags | Authority, Urgency, Secrecy, Fear, Social-Proof, Intimacy, Shame, Opportunity | All triggers identified; multi-value |
| decision_window | Immediate, Compressed, Same-Day, Multi-Day | Duration from STA contact to compliance |
| verification_attempt | Y, N, Partial | Did victim attempt out-of-band verification? |
| verification_breakdown | Channel-contaminated, No-OOB-check, Synthetic-callback, Suppressed, N/A | Why verification failed |
| requested_action | Fund-transfer, Credentials, MFA-code, Document, Account-reset, Other | Primary compliance action |
| ai_usage_confidence | Confirmed, Likely, Alleged | Based on source documentation quality |
| outcome_loss_usd | Numeric or Unknown | Reported financial loss in USD equivalent |
| outcome_completed | Y, N, Partial | Was compliance action completed? |

*Table 1. STAM Incident Coding Schema — 17 fields for cross-jurisdictional analysis.*

| Field | Coded Value (HK-2024-001) |
|---|---|
| incident_id | HK-2024-001 |
| date_range | January 2024 |
| jurisdiction | HK (primary); UK (impersonated origin) |
| victim_segment | Enterprise |
| impersonated_identity | Executive (CFO) and multiple colleagues |
| channel_sequence | Email, then video conference, then messaging |
| channel_switch_count | 2 |
| modality_set | Text, Audio, Video (Multimodal) |
| trigger_tags | Authority, Urgency, Secrecy, Social-Proof |
| decision_window | Compressed (same-day based on case narrative) |
| verification_attempt | N (not documented) |
| verification_breakdown | Channel-contaminated; Social-proof suppression |
| requested_action | Fund-transfer |
| ai_usage_confidence | Confirmed (police assessment) |
| outcome_loss_usd | Approximately 25,600,000 (HK$200M conversion) |
| outcome_completed | Y |

*Table 2. Pilot coding of the Hong Kong CFO deepfake case (January 2024) using the STAM schema.*

## 7. Testable Predictions and Hypotheses

The following four hypotheses are derived from the STAM framework and grounded in current evidence. They are stated in falsifiable form with explicit operationalizations, enabling future empirical testing using the STAM Incident Coding Schema or controlled experimental designs. Figure 3 visualizes the decision compression mechanism underlying H1.

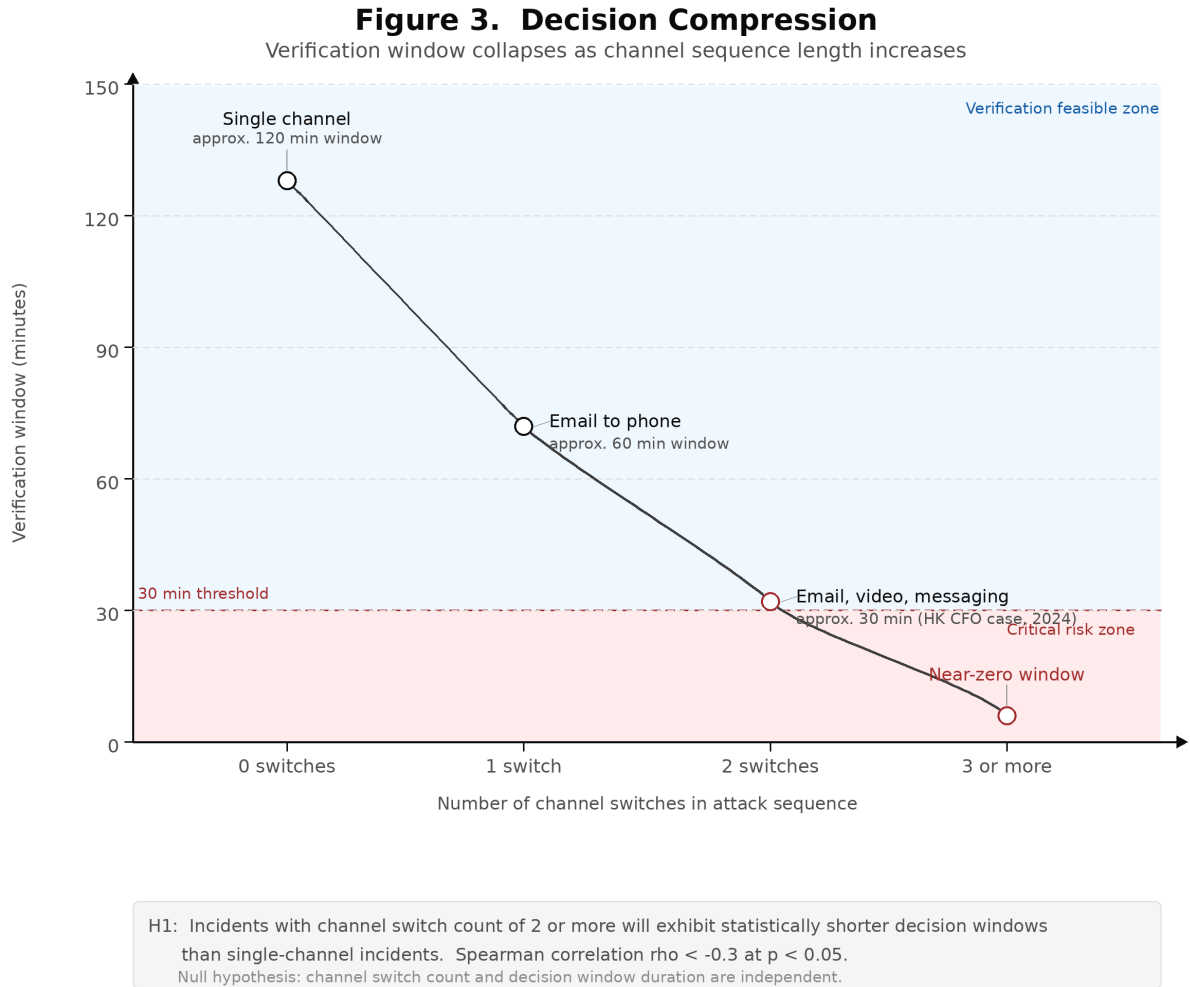

*Figure 3. Decision Compression. The verification window available to the victim collapses as the number of channel switches in the attack sequence increases. The Hong Kong CFO case (2 switches) sits at the 30-minute critical risk threshold. The H1 hypothesis predicts this relationship will be statistically confirmed in a coded incident corpus.*

### H1: Multi-Channel Orchestration Correlates with Shorter Decision Windows

Hypothesis: In a coded incident dataset, incidents with channel_switch_count of 2 or more will exhibit statistically shorter decision_window values than single-channel incidents, and this relationship will strengthen over time (2020 to 2026 temporal analysis). Null hypothesis: channel switch count and decision window duration are independent. Rejection criterion: Spearman rank correlation rho less than negative 0.3 at p less than 0.05.

## H2: LLM-Mediated Interactions Achieve Higher Compliance in Grooming-Phase Scams

Hypothesis: Among incidents coded as likely or confirmed AI involvement with decision_window of multi-day, compliance rates will exceed those in single-contact incidents, consistent with the LLM agent's demonstrated 46% versus 18% compliance advantage in extended conversations [5]. Null hypothesis: AI involvement has no effect on compliance rates across decision-window duration categories. Rejection criterion: Fisher's exact test at p less than 0.05.

## H3: Verification-Channel Attacks Will Increase as Callback Norms Spread

Hypothesis: As organizations adopt out-of-band callback verification protocols, adversaries will increasingly operationalize synthetic callback interception: answering verification calls with cloned voice personas or positioning synthetic agents on expected callback numbers. In the coded dataset, the proportion of incidents with verification_breakdown coded as synthetic-callback will increase significantly after 2024. Null hypothesis: verification_breakdown type distribution is stable across time. Rejection criterion: chi-square test at p less than 0.05.

## H4: Provenance Weaponization Will Grow as C2PA Adoption Increases

Hypothesis: As C2PA content credential adoption increases, adversaries will increasingly spoof, corrupt, or strategically absent provenance signals, exploiting the NIST-documented trust side effect whereby falsified provenance reduces trust even in authentic content [6]. The proportion of incidents with trust-cue category provenance coded as present will increase monotonically with C2PA adoption metrics. Null hypothesis: provenance-related trust cues are independent of C2PA adoption rates. Rejection criterion: Pearson correlation r greater than 0.5 at p less than 0.05.

# 8. Defensive Framework: Calm, Check, Confirm

The Calm, Check, Confirm framework, developed from practitioner experience and previously described in work deposited on Zenodo (ORCID: 0009-0001-3400-5016), is here operationalized as a research-grade defensive protocol with falsifiable components. The framework addresses the STAM Stage 6 mechanism directly: by introducing a mandatory cognitive pause and structured verification sequence, it targets decision compression as the primary attack vector.

## 8.1 Calm: Cognitive Interruption Protocol

The Calm component requires a mandatory pause before any high-value compliance action, regardless of urgency framing. Operationally, this is a minimum five-minute delay between receiving a high-value request and initiating any authorization sequence, explicitly overriding urgency claims from any channel. Dual-process theory predicts that this delay is sufficient to initiate deliberative System 2 processing even under urgency pressure [14]. Falsifiable prediction: organizations implementing a mandatory pre-authorization pause policy will show statistically lower rates of successful STA completion than matched organizations without such policies.

## 8.2 Check: Out-of-Band Verification

The Check component requires identity verification through a channel independent of those used in the initiating interaction. Verification must be initiated through a pre-stored, independently confirmed contact method, not a callback number or link provided within the interaction. This directly counters the channel contamination mechanism in STAM Stage 4. The FBI 2025 advisory independently recommends verification of AI-generated voice messages through known contact numbers [8].

### 8.3 Confirm: Two-Person Authorization

The Confirm component requires that any transaction above a defined threshold receive approval from two independently authorized individuals, each of whom must independently verify the request's legitimacy through out-of-band channels. The two-to-say-yes principle exploits the attacker's practical constraint: executing a full STA against two independent individuals simultaneously is operationally costly in a way that single-target attacks are not.

## 9. Discussion

### 9.1 Implications for Enterprise Security

The STAM framework reveals a fundamental misalignment in current enterprise security investment: the majority of resources are directed toward detection, while the attack has migrated to the decision layer. An organization with robust two-person authorization requirements and out-of-band verification protocols is more resistant to STAs than one with sophisticated media-detection tools but no pre-authorization pause policy.

### 9.2 Implications for Policymakers and Standards Bodies

The Trust-Cue Taxonomy provides a structured framework for evaluating proposed policy interventions including AI disclosure requirements, content watermarking mandates, and provenance certification schemes. Policymakers evaluating C2PA adoption should be informed by the NIST finding that falsified provenance information can reduce trust in authentic content [6]. Provenance standards must be paired with robust adversarial-resistance testing, not merely technical correctness verification.

### 9.3 Limitations

The current study has three primary limitations. First, the incident corpus is derived from public sources, introducing selection bias toward high-salience, high-loss incidents. Second, no controlled experiment has yet tested STAM variables in an ethically approved research design; causal claims are grounded in incident analysis and established behavioral science rather than direct experimental manipulation. Third, the pilot coding schema has been applied to a single incident; inter-rater reliability has not yet been established.

## 10. Future Work

Four research directions follow directly from this paper's contributions. First, the STAM Incident Coding Schema should be applied to a cross-jurisdictional corpus of at least 100 coded incidents with inter-rater reliability established, enabling the first statistically powered tests of H1 through H4. Priority sources include IC3 annual reports, INTERPOL and Europol operational summaries, and national law enforcement agency advisories.

Second, IRB-approved controlled experiments should operationalize STAM Stage 6 decision compression in laboratory and field settings, drawing on established social engineering research paradigms. Third, the Trust-Cue Taxonomy should be extended to capture emerging cue categories, with potential integration of neurological correlates of decision compression detectable through fMRI and EEG studies. Fourth, international regulatory frameworks informed by the STAM coding schema should be developed in collaboration with INTERPOL, UNODC, and national financial intelligence units.

## 11. My Final Conclusion

Generative AI has not invented a new category of crime. It has industrialized an ancient one. The capacity to manufacture trust has always been the core capability of skilled fraud operators. What generative AI provides is the ability to deploy that capacity at industrial scale, across thousands of targets simultaneously, with personalization that previously required months of human relationship-building.

This paper's central argument is that the appropriate response to this challenge is not primarily technical detection but architectural protection of the decision layer. The STAM framework provides the first formal, codeable model of the full Synthetic Trust Attack chain, enabling empirical analysis, policy evaluation, and defense design at the decision level rather than the perception level. The Trust-Cue Taxonomy provides structured vocabulary for cross-disciplinary analysis. The Incident Coding Schema enables reproducible empirical work. And the four testable hypotheses provide a research agenda that is falsifiable, not speculative.

Synthetic credibility, not synthetic media, is the true attack surface of the AI fraud era. Protecting it requires collaboration of computer scientists, behavioral psychologists, criminologists, policymakers, and practitioners. This paper is offered as a formal foundation for the empirical work that must follow.

## References

## AI Assistance Disclosure

In accordance with arXiv submission guidelines and emerging best practices for AI-assisted research, the author discloses that Claude (Anthropic, claude.ai, Sonnet 4.6) was used as an AI writing assistant during the preparation of this manuscript. Specific uses included structural organization of sections, language editing and readability improvement, figure generation support, and formatting assistance. All intellectual contributions are the original work of the author, including the STAM framework and its eight-stage model, the Trust-Cue Taxonomy, the Incident Coding Schema, the four falsifiable hypotheses (H1 through H4), the Calm, Check, Confirm defensive framework, the theoretical framing of decision compression as a formal attack variable, and all analysis, interpretation, and conclusions drawn. The author takes full responsibility for the scientific content, accuracy of citations, originality of contributions, and integrity of this work. AI tools were not used to generate data, fabricate results, or produce citations.

[1] Federal Bureau of Investigation, Internet Crime Complaint Center (IC3). Internet Crime Report 2024. U.S. Department of Justice, 2025.

[2] World Economic Forum. Global Cybersecurity Outlook 2025. WEF, Geneva, 2025.

[3] Europol Innovation Lab. Facing Reality? Law Enforcement and the Challenge of Deepfakes. Publications Office of the European Union, Luxembourg, 2022.

[4] INTERPOL. INTERPOL Cybercrime: Asia-Pacific Region Cyberthreat Assessment 2024. INTERPOL, Singapore, 2024.

[5] Gressel, G., Pankajakshan, R., Rozenfeld, S., Li, L., Franceschini, I., Achuthan, K., and Mirsky, Y. Love, Lies, and Language Models: Investigating AI's Role in Romance-Baiting Scams. arXiv:2512.16280 [cs.CR], December 2025. Available: https://arxiv.org/abs/2512.16280

[6] National Institute of Standards and Technology (NIST). Reducing Risks Posed by Synthetic Content: An Overview of Technical Approaches to Digital Content Transparency. NIST AI 100-4, U.S. Department of Commerce, 2024.

[7] Hong Kong Government. LCQ7: Combating Deepfake Fraud. Press Release, Legislative Council of Hong Kong, March 2024.

[8] Federal Bureau of Investigation. Public Service Announcement: FBI Warns of Increasing Threat of Cyber Criminals Utilizing Artificial Intelligence. FBI PSA, 2025.

[9] United Nations Office on Drugs and Crime (UNODC). Cybercrime and Fraud: A Rapidly Evolving Threat. UNODC Research Brief, Vienna, 2024.

[10] Kobis, N., et al. Fooled Twice: People Cannot Detect Deepfakes but Think They Can. iScience, 24(11), 2021.

[11] Bethany, M., et al. Generative AI in Social Engineering Attacks: A Systematic Review. ACM Computing Surveys, 57(3), 2024.

[12] Chesney, R. and Citron, D.K. Deep Fakes: A Looming Challenge for Privacy, Democracy, and National Security. California Law Review, 107(6), 2019.

[13] Coalition for Content Provenance and Authenticity (C2PA). C2PA Technical Specification Version 2.1. contentauthenticity.org, 2024.

[14] Cialdini, R.B. Influence: The Psychology of Persuasion. Harper Business, New York, 7th ed., 2021.

[15] Tolosana, R., et al. Deepfakes and Beyond: A Survey of Face Manipulation and Fake Detection. Information Fusion, 64, pp. 131-148, 2020.

[16] Yi, X., et al. Audio Deepfake Detection: A Survey. arXiv:2308.14970, 2023.

[17] Hancock, J.T., et al. AI-Mediated Communication: Definition, Research Agenda, and Ethical Considerations. Journal of Computer-Mediated Communication, 25(1), 2020.

[18] Hadnagy, C. Social Engineering: The Science of Human Hacking. Wiley, 2nd ed., 2018.

[19] Kahneman, D. Thinking, Fast and Slow. Farrar, Straus and Giroux, New York, 2011.

[20] Vaccari, C. and Chadwick, A. Deepfakes and Disinformation: Exploring the Impact of Synthetic Political Video on Deception, Uncertainty, and Trust in News. Social Media + Society, 6(1), 2020.

[21] Verdoliva, L. Media Forensics and Deepfakes: An Overview. IEEE Journal of Selected Topics in Signal Processing, 14(5), 2020.

[22] Masood, M., et al. Deepfakes Generation and Detection: State-of-the-Art, Open Challenges, Countermeasures, and Future Directions. Applied Intelligence, 53, 2023.

[23] FTC. FTC Warns of Impersonation Scams Using AI-Generated Voice. Federal Trade Commission Consumer Alert, 2023.

[24] Heartfield, R. and Loukas, G. A Taxonomy of Attacks and a Survey of Defence Mechanisms for Semantic Social Engineering Attacks. ACM Computing Surveys, 48(3), 2015.

[25] Anderson, R. Security Engineering: A Guide to Building Dependable Distributed Systems. Wiley, 3rd ed., 2020.

[26] Ashraf, M.T. (BeyondTahir). AI-Enabled Fraud and the Calm-Check-Confirm Defence Model. Zenodo Research Deposit, 2024. ORCID: 0009-0001-3400-5016.

[27] McAfee. The Artificial Imposter: AI-Powered Voice Cloning Study. McAfee Threat Research, 2023.

[28] Schmitt, M. and Flechais, I. Digital Deception: Generative Artificial Intelligence in Social Engineering and Phishing. Artificial Intelligence Review, Springer, 2024.

[29] INTERPOL. Facing the Reality of AI-Enhanced Fraud. INTERPOL Thematic Report, 2024.

[30] Luber, S., et al. The Economics of Cybercrime: Evidence from the Dark Web. American Economic Review, 114(2), 2024.

# Appendix A: STAM Codebook — Full Field Definitions

The following codebook provides operational definitions for all schema fields. Coders should apply the codebook independently and reconcile disagreements before finalizing codes. Inter-rater reliability should be assessed using Cohen's Kappa for categorical fields and intraclass correlation coefficient (ICC) for continuous fields.

| Field | Operational Definition | Coding Rule |
|---|---|---|
| incident_id | Unique alphanumeric identifier | Format: ISO-Country, YYYY, NNN |
| date_range | Time period during which the STA occurred | Use earliest and latest known incident dates |
| victim_segment | Organizational or demographic category of primary victim | Enterprise: professional capacity. Consumer: personal capacity. Government: public sector. |
| impersonated_identity | Category of trusted identity simulated by the adversary | Code the most senior identity; multi-value if multiple identities simulated simultaneously |
| channel_sequence | Ordered list of communication channels used in the attack | List channels in temporal order from first contact to compliance extraction |
| trigger_tags | Psychological compliance triggers identified in incident narrative | Code all present; minimum one required for inclusion |
| decision_window | Duration from first STA contact to compliance action | Immediate: under 5 minutes. Compressed: 5 to 30 minutes. Same-Day: 30 minutes to 24 hours. Multi-Day: over 24 hours. |
| verification_breakdown | Mechanism by which out-of-band verification failed or was not attempted | Channel-contaminated: verification in same channel family as attack. Synthetic-callback: verification call answered by synthetic persona. Suppressed: urgency framing prevented attempt. |
| ai_usage_confidence | Confidence level of AI involvement attribution | Confirmed: technical analysis or official attribution. Likely: synthetic media characteristics present. Alleged: reported without corroboration. |

## Appendix B: Practitioner Notes

Note: The following section represents the author's practitioner experience and design rationale. It is clearly labeled as practitioner perspective and should not be treated as empirical evidence for the main claims of this paper.

In the course of founding and operating PureDesigners Ltd., an AI and digital marketing agency serving clients in the UK, Qatar, and the United Arab Emirates, and in my role as Chair of the AAAI Pakistan Chapter (Chapter ID: 647885), I have encountered a consistent pattern across organizational fraud incidents: the moment of compliance is not a moment of ignorance. It is a moment of manufactured certainty.

Victims of the most sophisticated AI-enabled fraud incidents I have reviewed did not fail to recognize that fraud was possible. They failed to recognize that the specific interaction they were engaged in was the fraud. The attack succeeded not by deceiving their general knowledge but by overwhelming their specific situational awareness.

The Calm, Check, Confirm framework emerged from repeated observation of this pattern across geographies and organizational types. The three components are not novel security principles. What is novel is their framing as decision-layer defenses specifically calibrated against AI-synthesized trust attacks, and the recognition that they must function under the exact conditions STAs create: stress, authority pressure, time constraint, and social proof. Verification protocols that require calm deliberation to execute will fail under manufactured urgency. Effective defenses must be stress-resilient by design.

I offer this practitioner perspective not as scientific evidence but as design rationale, the experiential foundation from which the Calm, Check, Confirm framework was constructed, and which the empirical research agenda in this paper is designed to test.

---

*End of Paper*
Muhammad Tahir Ashraf (BeyondTahir) | ORCID: 0009-0001-3400-5016 | April 2026